# Which Generation Shows the Most Prudent Data Sharing Behaviour?

*Wolfgang Leister and Ingvar Tjøstheim*

*Norsk Regnesentral, Oslo, Norway.*

*Corresponding address: wolfgang.leister@nr.no*

## Abstract

We report from a study performed in ten European countries, where we asked about attitudes and behaviour towards data sharing behaviour. We looked into the differences between members of age groups. We find that there are more similarities than differences between the age groups, with the exception of young people more often tending to use fake information for privacy reasons. When analysing whether users change privacy settings as an indicator for awareness, we find that both the younger and the older users have lower awareness than the members of the middle-aged. The use of learning and practising tools seems the right way to increase the privacy and data sharing awareness of citizen.

## Introduction

The role of personal data and sharing in the digital economy is an important subject with implications for many areas of our societies. The discussion of the data sharing should be based on knowledge, both theoretical knowledge and empirical knowledge on what people actually do. When apps and digital appliances are personalised, data about the individual is used. The individuals behavior is collected and analyzed and often shared with third parties, but most people are unaware of this practise, and what happens with their data behind the scenes. Further, they are unable to predict the consequences of sharing and potential misuse.

When people are concerned about negative consequences from their data sharing, are they then prohibiting data sharing? The success of the social media suggests otherwise, and people seem to share quite much on social media (Durante 2011; Pingitore et al. 2017). Young people, that is the 16-19 year-olds, have a high willingness to learn. They absorb technology rapidly and use it in their everyday life. However, does this imply that they understand the dangers of sharing their data appropriately?

We explore this question in a survey in ten European countries supplied against the backdrop of a recent study about *Europeans' attitudes towards cyber security* (European Commission 2017) that included all European countries to find out about citizens' data sharing behaviour divided into different age groups. This research has as a goal to find out which generation has the most prudent behaviour to face the challenges that come with data sharing.

Humans do not always make choices that are in accordance with their own interests, as documented in the literature about choice behaviour (Kahneman 2011). Further, privacy is often difficult to understand for common people, as stated by the phrase "People don't read privacy policies; if they read, they wouldn't understand; if they understood, they wouldn't act." (Borgesius 2013).

## The ALerT Survey in Ten European Countries

To get evidence about citizens' behaviour and attitudes towards data sharing we performed a survey in ten European countries, here referred to as the ALerT survey[1]. In each of the countries Norway, Sweden, Denmark, Finland, and Germany we asked 100 persons in the age group 16-25 years and 100 persons in the age group 26-65 years. In France, Italy, The Netherlands, Poland, and the UK we asked 100 persons in total in the age group 16-65. In total, 1605 responses were recorded. These responses can be categorised by country, age, gender, and education.[2] To enable a comparison between all ten countries, we created a reduced data set of 1226 responses, where we adapted the age distribution of the countries with 200 responses to the age distribution of the remaining countries. To achieve this, we randomly removed responses in age groups that were over-represented in the data set of the five respective country.

---

[1] ALerT is the project name under which this research was funded.
[2] Due to the different education systems in the respective countries, and how the countries distinguish between primary, secondary and higher education, a directed comparison is not appropriate. Thus, we do not use education status further in our analysis.

Aside demographics, the questions in the survey were about self-assessment of own skills, attitudes, and remembered behaviour. The relevant questions are listed in Table 1. In Table 1, we omit several questions of our questionnaire, as these are not relevant here. Further, there was a freeform question about data sharing experiences, which we do not consider relevant for this paper.

We acknowledge that a survey only can record the stated attitudes, which may differ from the real behaviour by several reasons. There is a variety of biasing factors, and the differences we observed in the stated answers between the ten European countries suggests that cultural factors might play a role. To uncover whether the stated opinion is in accordance with the real behaviour, we need behavioral data from the respondents. Thus, other methods than surveys might be needed if one wants to study behaviour in addition to attitudes. However, we can assume that respondents can report on their own behavior in general, their mobile phone use, but we cannot assume that a person can remember exactly how many times the person has made changes in for instance the settings of the mobile phone.

Although there might be differences in the preferences throughout the ten European countries due to historical, cultural, and other reasons, our research sees the citizens in these countries as part of a digital society that is driven by an extensive due of digital services and digital applications. The legal framework is similar for the European countries. This was recently enforced with the advent of the GDPR (European Parliament 2016) that came into force in May 2018.

We present the results of our study where we present the results of ten questions of relevance (see Table 1). We show the results of each of three age groups for all ten European countries, in addition to the averages for each of the age groups. We assume that the attitude and behaviour have a similar pattern across the countries. Usually, if a majority of the participants are on the positive side, i.e., fully agree + agree, in most cases there will be a majority on the positive side also in the other countries.

Table 1: Questions (relevant questions selected)

| Reference | Question | Purpose |
|---|---|---|
| Q6 | Consider your computer or Internet skills. Do you know how to protect your personal data? | Skills (self-assessment) |
| Q12 | Do you agree or disagree: I have not done anything in particular to get more knowledge about data-sharing, privacy and how my personal data are used because I do not feel I need to? | Attitude |
| Q13 | About apps on smartphones. To be informed, I read the information under terms of use. | Attitude |
| Q14 | About apps on smartphones: If I am given the choice, I change the setting to reduce sharing of my data. | Attitude |
| Q15 | Do you agree or disagree: To get more knowledge about how personal data are used, I have asked questions in a group or forum, or asked someone I know to learn about privacy and protection of personal data. | Attitude |
| Q17 | Do you agree or disagree: An app-provider should be allowed to use your contact-list also for other purposes than the app needs to function. | Attitude |
| Q8 | Have you changed the privacy setting in your Internet-browser or an app to avoid sharing of personal data? | Behaviour |
| Q9 | When using or installing an app on your smartphone, have you restricted or refused access to your personal data (e.g., your location, contact list)? | Behaviour |
| Q10 | Have you filled in incorrect or fake information about yourself to a website or an app for privacy reasons? | Behaviour |
| Q16 | Have you decided not to download an app on your mobile phone because the app required personal information that you did not want to share (example: your contact-list). | Behaviour |

## Results from the ALerT survey

In our survey, we divided the responses into three age groups; 16-29 years, 30-49 years, and 50-65 years of age. Further, we divided the responses into the ten countries represented in the survey.

**Skills**

We used additionally the results from the European survey and some findings from *Europeans' attitudes towards cyber security (European Commission 2017)*. This is a large-scale survey (personal interviews with 22.236 respondents /Internet users), representing 340 mill. Europeans in 28 EU countries. We display findings from both surveys next to each other, although they use different sampling and interview methods. Figure 1 shows that 45% of European Internet-users are concerned about misuse of personal data. About half of the population answered that they are not well informed about the risk of cyber-crime, which may involve personal data and identity theft. In contrast to the EU-survey, the ALerT survey includes questions about concrete privacy protection practices.

In these studies 45% of European Internet-users are concerned about misuse of personal data. About half of the population answered that they are not well informed about the risk of cyber-crime, which may involve personal data and identity theft.

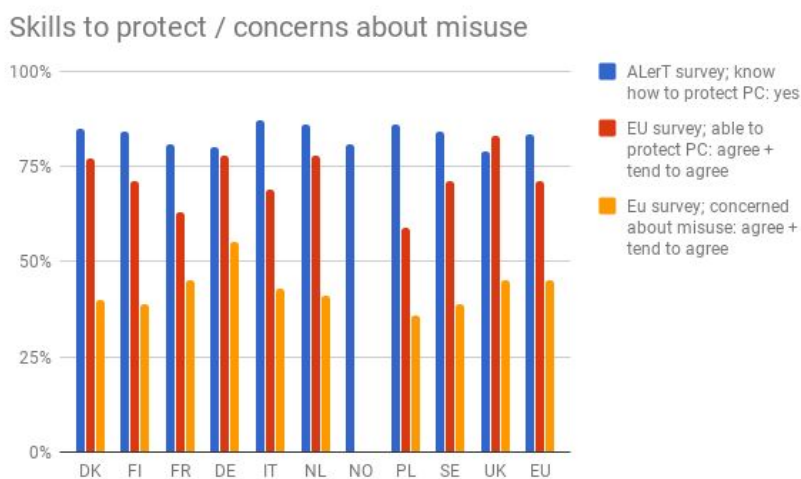

Figure 1: ICT knowledge and concerns about misuse of personal data among European citizens[3]

Do citizens have the necessary knowledge and skills to protect themselves? A survey cannot measure actual skills and people tend to overestimate their knowledge skills, as a recent review article states: *People may be overconfident in their assessment of privacy or security risks. For example, they may be convinced that their antivirus software is fully effective against all possible threats, while its efficacy may be substantially lower. As a consequence, they engage in insecure online behaviors that lead to compromised devices and data loss* (Acquisti et al. 2017). While we acknowledge the tendency to be overconfident, we need information about how people perceive their own skills. In both surveys, a large majority of the citizens reported that they know how to protect themselves.

We asked the participants in the survey to consider their skills to protect their personal data in Q6 (see Figure 2). Surprisingly a large majority answered that they have this skill (about 83% in average) with small differences among the age groups. Given this high number associated with the stated behaviour (the answers to the following questions in the survey), we have reasons to assume that the participants overestimated their own skills. This is in-line with related research (Acquisti et al. 2017).

**Attitude**

The results about the questions related to attitude are shown in Figure 2.

For the question on whether the person felt that she or he needed to get more knowledge on data-sharing (Q12) the 10 countries together, 45% of the youngest (age 15-29) agreed to this statement while 41% and 40% for the other age-groups. For the ten countries the numbers varied between 31%

---

(Germany) and up 55% (Italy). Our interpretation of this outcome suggests that a high agreement-percentage indicates a low privacy-concern.

We categorized the question whether the participants read the terms of use (Q13) as an attitude-question. This question was formulated as a general question that did not take into account how often or how many pages of the terms of use one reads. For age 16-29, the agreement ranged from 44% (Finland) to 86% (Germany). The average was 59%. Germany had the highest percentage, indicating a high concern for privacy. However, the interpretation of what is meant by "I read ..." might differ. For the two other age groups, the average percentages were higher with 70% for 30-49, and 73% for 50-65.

Statement Q14 addressed apps on smartphones and whether the respondent have changed settings to reduce sharing of data. In most cases, it is possible to change the privacy settings.[4] The contingency "if I am given the choice", was used because we are primarily interested in the attitude rather than the actual behaviour. The average percentage answering fully agree or agree was 81% for the age 16-29. The lowest was France with 60% and the highest the Netherlands with 93%. For the two other age groups, the average percentages were also high with 83% for 30-49, and 73% for 50-65.

Digital technologies constantly evolves, which makes it necessary for participants of digital services to renew their knowledge. Q15 was about whether the participants have asked questions in a forum or acquaintances on how to protect their data. On average 37% fully agreed or agreed, ranging from 30% (Italy) to 50% (UK) for age 16-29. Although knowledge about this subject can be gained by other means, such as reading articles in journals or news feeds, we interpreted these answers that there is a lack of active approach that knowledge needs constantly to be updated. For the two other age groups, the average percentages were on the same level with 40% for age 30-49, and 30% for age 50-65.

For all age groups (average) only about 11% (7%-14%) reported the response alternative "fully agree" to whether they have asked others about how to protect themselves. For 16-29 the percentage was 9%, for 30-49 the percentage was 15% and for 50-65 the percentage was 9%. This seems to contradict the finding about the self assessment of their skills. We assumed that it would be much more common to ask someone more knowledgeable about such issues, even for people with good skills.

For apps the settings might differ. Quite often, an app asks for permission to access the contact list on the phone. For an emergency app, this makes good sense, for instance to be able to contact your closest family members. For other apps it might be hard to find a reason why the app needs access to the contact list. Q17 reflected this by asking whether an app-provider should be allowed to use the contact list for other purposes than necessary for the app to function. For this question, we counted how many participants fully disagreed or disagreed whether the app provider should be allowed to access the contact list. For age 16-29 on average 63% disagreed; the highest was Norway with 78% and the lowest were the UK and France, both with 44%. Overall, the answers reflected a negative attitude towards this type of use of contact lists, and is, hence, a privacy-oriented attitude. For the two other age groups, the average percentages were higher with 75% for 30-49, and 71% for 50-65.

In summary, the answers to these five attitude questions indicated privacy concerns, precautions and attitudes in accordance to the norm "take responsibility" of your data. We observed some examples of differences between the age groups. These differences were mostly in favour of the age group 30-49, i.e., they were most concerned about data sharing in their attitude.

**Behaviour**
The results about the questions related to attitude are shown in Figure 3.

The next set of questions is about behaviour. There is no one-to-one relationship between what people answer in a survey and what people actually do (MacKenzie and Podsakoff 2012; LaPiere 1934). Instead of asking "do you agree to a statement", we asked whether the respondents have performed certain actions, and we assumed that this reflects behaviour or is intended to reflect behavior. We acknowledge that it sometimes may be difficult to remember a specific case and report accurately (Pryor et al. 1977), but it should be reasonable to remember whether settings on a smartphone have been actively changed.

---

[4] On all newer Android phones privacy and data sharing settings are available, while these settings are not always available on other architectures. Settings can be changed using the ask-on-install (AOI) model or ask-on-first-use model (AOFU), depending on the version of the operating system (Tsai et al. 2017).

Q8 was about whether the participant has changed the privacy settings in an Internet browser or an app to avoid sharing of personal data, 79% in the age group 16-29 have done this once or more. The percentages varied from 70% (The Netherlands) to 92% (France). For the two other age groups the average numbers were 83% for age 30-49, and 73% for age 50-65.

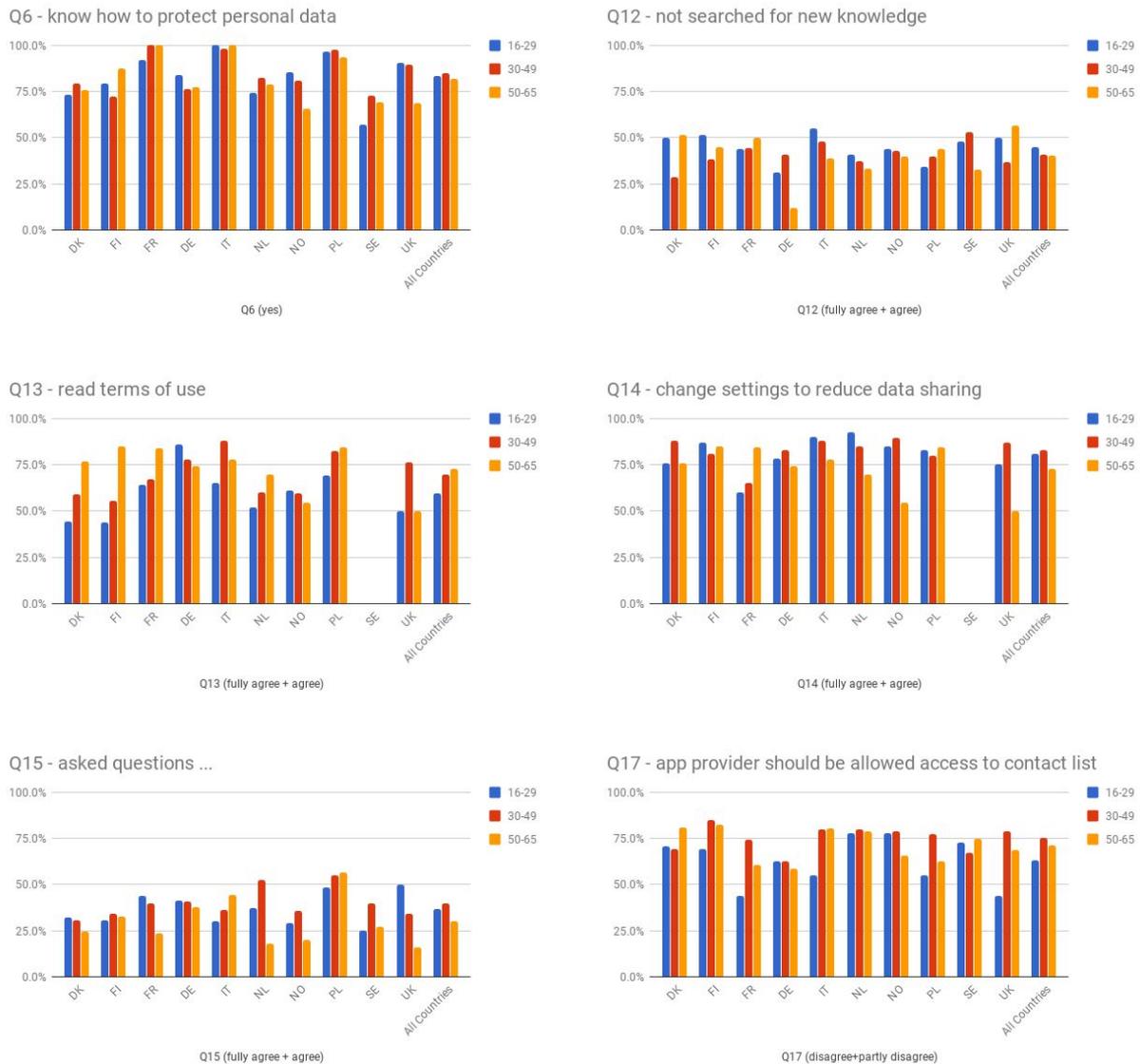

Figure 2: Skills (Q12) and attitudes (Q13, Q14, Q15, Q17) by country and age group

Q9 regarded whether the participants have restricted or refused access to personal data on a smartphone when using or installing an app. For age 16-29, 78% answered positively to that while the numbers were 75% for age 30-49 and 55% for age 50-65.

Many are aware of the fact that companies collect and store a lot of personal information. Given that a person disagrees to this practise, there is a variety of counter-steps that can be taken. Q10 asked whether the participants have filled in incorrect or fake information about themselves to a website or an app for privacy reasons. For the age 16-29, 57% answered that they have done this at least once. The lowest was 35% (Italy), and the highest is 84% (France). 46% of the age group 30-49 and only 23% in the age group

50-65 have done this. Thus, it seems that the youngest have a preference for this counter-step as compared to the older participants.

In some cases, changes to the privacy settings are not possible or too complex to perform. To protect their data, one can refuse to install the app or uninstall it, thus not being able to use this functionality. Q16 asked whether the participants have decided not to download an app on their mobile phone because the app required personal information that the participant did not want to share. 75% in the age group 16-29 did this once or more. Sweden had the lowest percentage of 65% and Finland the highest with 92%. For the two other age groups the average numbers were 78% for age 30-49, and 73% for age 50-65. The answers indicated that a majority knew that this might be the best alternative, but we assumed that this is based on a trade off that there are other better apps, or the functionality of the app in question was not really needed.

Overall, there were more similarities than differences between the age groups. Only for the question whether they have filled in fake information, the younger behave differently; they prefered this method to protect their privacy much more often than the older participants.

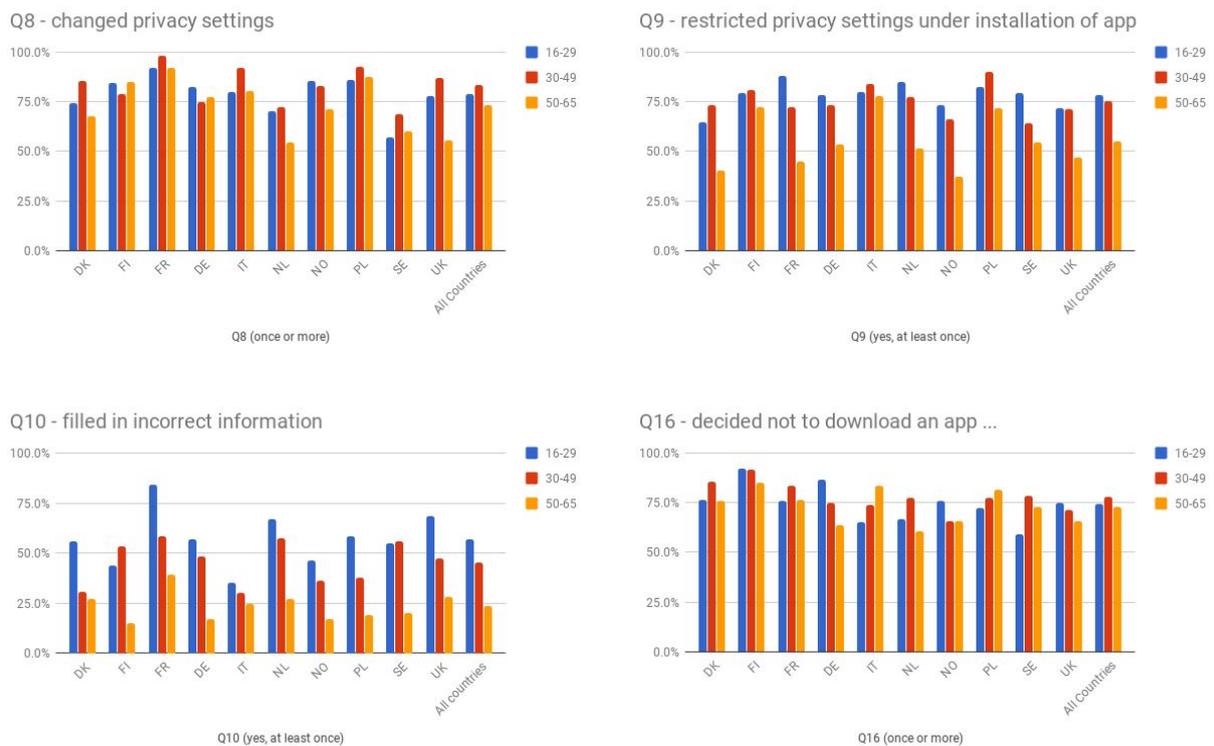

Figure 3: Behaviour (Q8, Q9, Q10, Q16).

### Awareness to protect personal data across generations

To take a closer look at the awareness to protect personal data across generations, we used the answers given in our survey (n=1605) and divided these into seven age groups: 16-19, 20-25, 26-29, 30-39, 40-49, 50-59, and 60-65. To describe awareness, we used a function based on the answers from Q8 and Q9, i.e., whether the participants had made changes in the privacy settings in their browser and whether they have made changes in the privacy settings when having installed an app. Respondents who answered "more than once" on both questions are classified into the high awareness group. Those answering that they did not make changes on both questions are classified into the low awareness group, while the remaining are classified into the medium awareness group. In what follows, we compare the high and low awareness groups.

In the high-awareness group, the 16-19 year-olds scored low with about 36%, and only the generation 60+ scored lower with 29%. In the other age groups, we see the trend that awareness rose to 52% for the 30-40 year-olds, before it decreased with higher age. For the low-awareness group, we observed the opposite behaviour: only 16% had low awareness, while this group was larger for both the younger and the older. This is shown in the graph in Figure 4. With reference to this figure, we can argue that in all groups privacy awareness should be increased, and more so for the youngest that had the lowest percentage. Only the 30-39 year olds were above 50%.

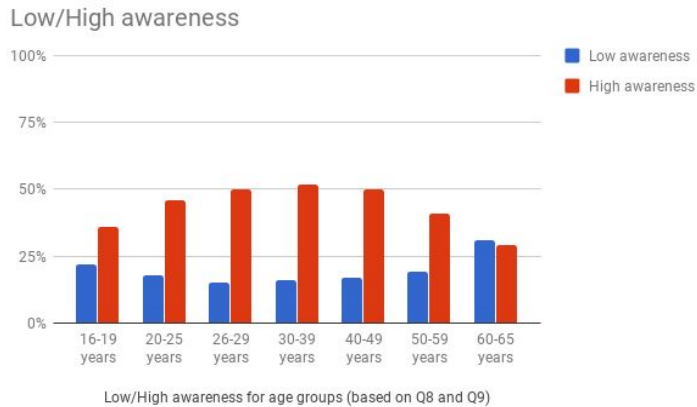

Figure 4: high awareness and low awareness among seven age groups

Since the younger generation seemed to have embraced technology while the older struggle with it, the fact that the 16-19 year-olds showed low awareness seems surprising. However, as the younger generation embraces games, so-called serious games, sometimes referred to as learning games, we envision that such serious games can be used to increase the awareness, which also would have a positive effect on the awareness in the population, as the now younger generation grows older. Using serious games has the potential to provide feedback to users, following the citation: "*The best way to help humans improve their performance is to provide feedback. Well-designed systems tell people when they are doing well and when they are making mistakes.*" (Thaler, Sunstein, and Balz 2010).

We conclude from these findings that significantly more than one third of the population needs help to tackle the challenges with sharing data everywhere, particularly the young (16-19) and the older (50+). The responses from our survey suggest that the 30-49 year-olds show the most prudent data sharing behaviour, which might be a European phenomenon, as an American study suggests that younger consumers take more protective actions (Pingitore et al. 2017).

## Conclusion

When using the Internet and apps, people need to make decisions about privacy and data sharing settings although these are not the main concern of the user. In most cases, these settings are hidden and are changed only seldomly. As a consequence, these settings get out of the focus for most users. Further, these settings are often presented in such a manner that users accept the default, as shown in a recent report (Moen, Ravna, and Myrstad 2018). This practise is referred to as privacy dark patterns (Bösch et al. 2016) and is based on architectural design patterns (Alexander et al. 1977). On-line services sometimes ask for additional personal data that is not necessary to interact with the service. This sweet-seduction approach (Fritsch 2017) is also a dark-pattern that is quite common.

People do not always make choices in accordance with their own self-interests. The reasons for the mismatch between attitudes (stated preferences) and revealed preferences are well-documented in the literature on choice behavior. In our study, we have used surveys to document how citizens perceive problems with sharing personal data, and steps taken to protect personal data.

To help citizens, we propose the use of serious games to assist humans without imposing overly prescriptive models of what the "right" decisions might be; there is a need for both assisting and practising tools. There are already some recent examples for such tools in the recent research literature (Tsai et al. 2017).

We have drawn attention to the 16-19 year-olds, as the willingness to learn is higher at an early age. According to our survey, we observed low awareness for them, but higher awareness and less sharing of personal data among the 20-25 year-olds, and even more for the 30-39 years old, that is, they take more control of their personal data. If we can increase the awareness among the young, this will have an impact on the privacy and data sharing awareness as the young generation grows older.

## Acknowledgement

This research is done in the context of the ALerT (Awareness Learning Tools for Data Sharing Everywhere) project funded by the Research Council of Norway, project number 270969.